\documentclass[showpacs, reprint,
nofootinbib,
amsmath, amssymb,
aps,
]{revtex4-1}

\usepackage{graphicx}
\usepackage{dcolumn}
\usepackage{bm}
\usepackage{hyperref}

\begin{document}

\preprint{APS/123-QED}

\title{$f(R)$-Modified Gravity, Wald Entropy, and the Generalized Uncertainty Principle}

\author{Fay\c{c}al Hammad}
\email{fhammad@ubishops.ca}
\affiliation{Physics Department \& STAR Research Cluster, Bishop's University, and
\\Physics Department, Champlain College-Lennoxville\\
2600 College Street, Sherbrooke, Qu\'{e}bec J1M 1Z7, Canada}


\begin{abstract}
Wald's entropy formula allows one to find the entropy of black holes' event horizon within any diffeomorphism invariant theory of gravity. When applied to general relativity, the formula yields the Bekenstein-Hawking result but, for any other gravitational action that departs from the Hilbert action, the resulting entropy acquires an additional multiplicative factor that depends on the global geometry of the background spacetime. On the other hand, the generalized uncertainty principle (GUP) has extensively been recently used to investigate corrections to the Bekenstein-Hawking entropy formula, with the conclusion that the latter always comes multiplied by a factor that depends on the area of the event horizon. We show, by considering the case of an $f(R)$-modified gravity, that the usual black hole entropy derivation based on the GUP might be modified in such a way that the two methods yield the same corrections to Bekenstein-Hawking formula. The procedure turns out to be an interesting method for seeking modified gravity theories. Two different versions of the GUP are used and it is found that only one of them yields a viable modified gravity model. Conversely, it is possible to find a general formulation of the GUP that would reproduce Wald entropy formula for any $f(R)$-theory of gravity.
\end{abstract}

\pacs {04.70.Dy, 11.30.-j, 04.50.-h}
\maketitle

\section{Introduction}\label{sec:1}
Modified gravity is a fertile area of research when it comes to tackling problems from cosmology \cite{Capozziello1, Felice, Nojiri1, Capozziello2} or investigating the problem of quantum gravity \cite{Olmo}. However, in the absence of any guiding principle, one a priori 'invents' plausible models and then confronts them with observation. Therefore, any experimental or theoretical hint that might favor one model over another could play a major role in finding the right modifications one should bring to the Hilbert gravitational action to answer some of the longstanding problems in gravity.

One of the procedures used in selecting one modified gravity from another is to study the same phenomenon within different models and then adjust, or modify, the more flexible of these models to achieve compatibility of their respective predictions. This is the strategy that we shall follow in this paper. In fact, such strategy has recently been used in Refs.~\cite{Oikonomou, Odintsov} in the search for $f(R)$ and $f(T)$ modified gravity theories by looking for those theories that would reproduce the cosmological bouncing scenario as it already arises within Loop Quantum Cosmology \cite{Ashtekar} and the Ekpyrotic scenario \cite{Khoury}.

Another approach in searching for modified gravity theories consists in imposing fundamental principles from thermodynamics on the general structure of a model in order to come up with constraints to be imposed on that structure and, hence, select among the huge possibilities one could a priori include inside the original model. One such approach has recently been pursued in Ref.~\cite{Hammad1} and yielded non-trivial constraints on possible modified gravity theories. This use of thermodynamics in the search for modified gravity is amply justified by the intriguing fact that thermodynamics might intimately be linked with the physics of spacetime. This is most apparent in the work of Ref.~\cite{Jacobson} where general relativity was derived from thermodynamics. Another case for this has been provided by Wald who showed in Ref.~\cite{Wald1} (see also \cite{Wald2}), by deriving a general entropy formula, that black holes' entropy comes from the Noether charge associated to any diffeomorphism invariant theory of gravity.

Wald's formula allows one to compute the entropy of an event horizon within any well-defined Lagrangian theory of gravity. When applied to the Lagrangian of general relativity \cite{Wald1, Wald2}, the formula gives exactly the Bekenstein-Hawking entropy $A/4G$ \footnote{The natural units $k_{B}=\hbar=c=1$ will be used throughout this paper, where $k_{B}$ is the Boltzmann constant.}, where $A$ is the area of the event horizon and $G$ is Newton's gravitational constant. When applied to a gravitational action that departs from the Hilbert action, however, Wald's formula still produces an entropy which is proportional to the area of the event horizon but in which the proportionality factor depends on the curvature of the background spacetime \cite{Vollick, Briscese, Cruz1, Cruz2}. Therefore, if it happens that a different fundamental approach also yields departures from the Bekenstein-Hawking formula, one might profitably use that approach to 'calibrate' one's favorite modified gravity model using Wald's entropy formula.

One such different fundamental approach is based on the so-called generalized uncertainty principle (GUP) which came from various approaches to quantum gravity as a generalization of Heisenberg's uncertainty principle of quantum mechanics \cite{Mead, Veneziano, Amati, Gross, Yonega, Konishi, Guida, Luis}. In the usual Heisenberg's uncertainty principle, the product of the uncertainties on position and momentum has a minimum bound equal to Planck's constant $\hbar$. In the framework of the GUP, this minimum bound becomes increased by linear and quadratic terms in the uncertainty on momentum. These additional terms are responsible for bringing correction terms to the Bekenstein-Hawking formula when the GUP is used to investigate the near horizon thermodynamics \cite{Medved}.

The usual procedure \cite{Medved} used to extract black hole's entropy from the GUP gives an entropy that is proportional to the area of the event horizon but with a proportionality factor that is also a functional of the area $A$ of the event horizon. Therefore, the proportionality factor would be different for different black holes. As we mentioned above, however, modified gravity would yield via Wald's formula a unique proportionality factor for every black hole since it does not depend on the geometry of the event horizon but depends only on the global geometry of spacetime.

Now, since both the GUP-induced entropy and Wald entropy are based on fundamental arguments, they should in principle yield the same result when applied to the investigation of the same physical system. The fact that their respective results for black holes differ, means that either the deviations from the pure area-law that arise from the GUP are not due to corrections in general relativity, but simply result from higher-order contributions of vacuum fluctuations; or else, that the usual GUP-derivation of black holes' entropy could be modified in such a way that the result coincides with what modified gravity predicts through Wald's entropy. In the present work, we shall contemplate the latter possibility and use it as a tool to find modified gravity theories. The modified gravity theories we study in this paper will be the simple class of $f(R)$-gravity models, but the approach might readily be generalized to investigate other classes of more complicated modified gravity theories.

On the other hand, the fact that it is possible for the GUP to reproduce an $f(R)$-gravity, because it implies the same result coming from Wald's entropy, leads one to suspect that there might exist some general formulation of the GUP from which the fundamental Wald entropy formula itself might arise. We shall investigate this possibility by looking for such a formulation within the simple case of $f(R)$-gravity.

This paper is organized as follows. In Sec.~\ref{sec:2}, we give a brief introduction to the Noether charge method for computing black holes' entropy; we give the general Wald entropy formula, and we explicitly write down the result one obtains for the case of an $f(R)$-gravity. We then introduce the simplest version of the GUP formula and briefly explain the main steps one usually follows to obtain the entropy of black holes. In Sec.~\ref{sec:3}, we confront the black hole entropy coming from $f(R)$-gravity through Wald's formula with that obtained from the GUP. We argue for the necessity to slightly modify the usual procedure followed to obtain black holes' entropy from the GUP, and we give the entropy that results from this modification within an $f(R)$-gravity. Conversely, we show that in order for Wald's formula to yield exactly that same entropy as the one obtained from the GUP procedure thus modified, a specific model of $f(R)$-gravity is required. We show that in the simplest version of the GUP the $f(R)$-gravity that results is not viable but when using a more complicated version of the GUP, more recently introduced in the literature, one finds a viable model. In Sec.~\ref{sec:4}, we find a general formulation of the GUP from which Wald entropy arises. We end this paper with a brief conclusion and discussion.

\section{Wald Entropy vs. GUP-Induced Entropy}\label{sec:2}
A diffeomorphism invariant theory of gravity is a theory whose Lagrangian $\mathcal{L}(\phi)$, where $\phi$ denotes all the field variables on which the theory depends, is invariant under arbitrary spacetime transformations. Then, just as for any other theory with a given invariance, one can extract a Noether charge $\textbf{Q}$: a two-form associated with the Noether three-form current $\textbf{J}=d\textbf{Q}$ generated by the diffeomorphism transformations. As explained in Refs.~\cite{Wald1, Wald2}, since the variation $\delta\textbf{L}=\textbf{E}\delta\phi+d\mathbf{\Theta}(\phi,\delta\phi)$ of the four-form $\textbf{L}=\mathcal{L}(\phi)\mathrm{d}x^{0}\wedge...\wedge\mathrm{d}x^{3}$ gives the equations of motion $\textbf{E}$ (in form notation) up to a total divergence $d\mathbf{\Theta}$, where $\mathbf{\Theta}$ is a three-form, it follows that the Noether current is given by $\textbf{J}=\mathbf{\Theta}(\phi,\pounds_{\xi}\phi)-\xi.\textbf{L}=d\textbf{Q}$, where $\xi=\xi^{\mu}\partial_{\mu}$ is the vector field that generates the diffeomorphism and $\pounds_{\xi}$ is the Lie derivative with respect to $\xi$. It was shown in \cite{Wald1, Wald2} that the flux of this current through the bifurcation surface $\mathcal{H}$ of a black hole horizon is nothing but the entropy transfer between the two regions of spacetime separated by that surface. That entropy is called the Wald entropy and is defined by $S=(2\pi/\kappa)\int_{\mathcal{H}}\textbf{Q}$, where $\kappa$ is the surface gravity of the horizon. For a gravitational theory whose Lagrangian is $\mathcal{L}(g_{\mu\nu}, R_{\mu\nu\rho\sigma},\nabla_{\mu}R_{\nu\rho\sigma\lambda}, ...)$, the entropy that would result for an event horizon $\mathcal{H}$ is given by \cite{Wald1, Wald2}:
\begin{equation}\label{1}
S=-2\pi\int_{\mathcal{H}}\sqrt{h}\,\mathrm{d}^{3}x\frac{\delta\mathcal{L}}{\delta R_{\mu\nu\rho\sigma}}\mathbf{\epsilon}_{\mu\nu}\mathbf{\epsilon}_{\rho\sigma},
\end{equation}
where $h$ is the induced metric on $\mathcal{H}$ whose binormal is $\epsilon_{\mu\nu}$ such that $\epsilon_{\mu\nu}\epsilon^{\mu\nu}=-2$, and the functional derivative of the Lagrangian is performed with respect to the Riemann tensor $R_{\mu\nu\rho\sigma}$ by keeping the metric $g_{\mu\nu}$ and the connection $\Gamma^{\rho}_{\mu\nu}$ variables fixed. In the special case of an $f(R)$-gravity theory, that is a theory whose gravitational action $S=(1/16\pi G)\int\mathrm{d}^{4}x\sqrt{-g}f(R)$ depends only on the Ricci scalar $R$ through the an \textit{a priori} arbitrary functional $f(R)$, the above formula gives \cite{Vollick, Briscese}
\begin{equation}\label{2}
S=\frac{A}{4G}\left[\frac{\partial f(R)}{\partial R}\right]_{\mathcal{H}},
\end{equation}
where the subscript $\mathcal{H}$ means that the quantity is evaluated on the event horizon $\mathcal{H}$.

Let us now apply the latter formula for the case of an event horizon due to a black hole in an $f(R)$-theory of gravity. For that purpose we need to evaluate the differential $\partial f/\partial R$ on the event horizon $\mathcal{H}$. Recall that the metric equations of motion in an $f(R)$-modified gravity in vacuum read $R_{\mu\nu}\partial f/\partial R-fg_{\mu\nu}/2=0$, the contraction of which gives $R\partial f(R)/\partial R=2f(R)$. Thus, unless $f(R)\propto R^{2}$, this equation implies that the Ricci scalar $R$ should be a constant \cite{Briscese, Cruz1, Cruz2}. Let us denote the constant curvature by $R_{0}$. Therefore, since $R$ would be equal in this case to $R_{0}$ everywhere, formula (\ref{2}) reads in an $f(R)$-gravity as
\begin{equation}\label{3}
S=\frac{A}{4G}\left[\frac{\partial f(R)}{R}\right]_{R=R_{0}}.
\end{equation}
The important feature to notice in this formula is that the content inside the square brackets on the right-hand side does not depend on the particular black hole of interest but is really a constant proper to the Universe that contains the black hole.

Let us now examine the GUP and its implications for black holes thermodynamics. What is meant by GUP (in its simplest version) is the following inequality satisfied by the product of the uncertainty on position $\Delta x$ and the uncertainty on momentum $\Delta p$ of a quantum particle \footnote{Here we restrain ourselves in this paper to the one-dimensional version of the GUP which is sufficient for the derivation of black holes thermodynamics.} \cite{Medved}:
\begin{equation}\label{4}
\Delta x\Delta p\geq1+\alpha^{2}l^{2}_{P}\Delta p^{2},
\end{equation}
where $\alpha$ is a dimensionless constant of order unity that depends on the mathematical model used to investigate the physics at the Planck scale, and $l_{P}$ is just the fundamental Planck length. Recall, as we shall make use of it below, that in terms of Newton's gravitational constant $G$, we have in our units the following identity $l^{2}_{P}=G$.

The way inequality (\ref{4}) is used to obtain the entropy of a black hole whose event horizon has area $A$ consists in the following five steps \cite{Medved}. First, from inequality (\ref{4}) one finds a lower bound for the uncertainty on momentum $\Delta p$ in terms of the uncertainty on position $\Delta x$ as follows: $2\alpha^{2}l^{2}_{P}\Delta p\geq\Delta x-\sqrt{\Delta x^{2}-4\alpha^{2}l^{2}_{P}}$, where the minus sign outside the square root has been chosen so that one recovers the Heisenberg uncertainty relation when making $l_{P}\rightarrow0$. Second, the uncertainty $\Delta p$ in the latter inequality is traded for the uncertainty on energy $E$, just as it is done within the conventional Heisenberg uncertainty principle where one extracts from $\Delta p\geq1/\Delta x$ the lower bound for energy $E\geq1/\Delta x$. By multiplying then both sides of the resulting inequality by $\Delta x$, one finds $2\alpha^{2}l^{2}_{P}E\Delta x\geq\Delta x^{2}-\Delta x\sqrt{\Delta x^{2}-4\alpha^{2}l^{2}_{P}}$. Third, by combining this inequality with the fact that the minimum increase $\Delta A_{min}$ of a black hole horizon that absorbs a quantum of energy $E$ that has a spatial extension (or 'size') $\Delta x$ is given by $\Delta A_{min}\geq 8\pi l^{2}_{P}E\Delta x$, one finds that
\begin{equation}\label{5}
\Delta A_{min}\geq\frac{4\pi}{\alpha^{2}}\left(\Delta x^{2}-\Delta x\sqrt{\Delta x^{2}-4\alpha^{2}l^{2}_{P}}\right).
\end{equation}

Now comes the fourth and crucial step in relating the GUP to the entropy of black holes. One usually assigns to the uncertainty on position $\Delta x$ of the quantum particle absorbed by the black hole the following estimate: $\Delta x\sim 2r_{S}=\sqrt{A/\pi}$, where $r_{S}$ is the Schwarzschild radius. This choice has been justified in \cite{Medved} by arguing, on one hand, that in a microcanonical framework, a black hole should be viewed as a macroscopically large system of fixed mass and in thermodynamical equilibrium with the bath of radiation -- the Hawking radiation in this case -- whose temperature is the Hawking temperature as seen by an asymptotic observer. Therefore, the size of the particles composing the bath is of the order of their Compton wavelength, which is the inverse of the Hawking temperature. On the other hand, the choice $\Delta x\sim 2r_{S}$ also appears as the natural scale in the near-horizon geometry since it is of the order of the inverse of the surface gravity $\kappa^{-1}=2r_{S}$ \cite{Medved}.

The fifth and last step that leads to the entropy of the black hole is the assumption that, since any area increase in the event horizon should yield an increase in entropy, the elementary increase $\Delta S$ in entropy of the black hole due to the elementary increase $\Delta A_{min}$ in the area should be one bit of information, i.e. one fundamental unit of entropy, denoted $b$. Using this together with the above choice for $\Delta x$ inside inequality (\ref{5}) leads to the following differential equation for entropy \cite{Medved, Awad}:
\begin{equation}\label{6}
\frac{\mathrm{d}S}{\mathrm{d}A}=\frac{b}{\Delta A_{min}}=\frac{\alpha^{2}b}{4\epsilon\left(A-\sqrt{A^{2}-4\pi\alpha^{2}GA}\right)},
\end{equation}
where we have introduced Newton's gravitational constant inside the square root and introduced the numerical factor $\epsilon$ that would fixe the specific value of the minimum increase $\Delta A_{min}$ in area \cite{Medved}. When integrated, Eq.~(\ref{6}) gives, up to an integration constant, the entropy of the black hole \cite{Awad}:
\begin{multline}\label{7}
S=\frac{bA}{16\pi\epsilon G}\times\\
\left[1+\sqrt{1-\frac{\beta}{A}}-\frac{\beta}{2A}\ln\left(A+\sqrt{A^{2}-\beta A}-\frac{\beta}{2}\right)\right]
\end{multline}
where we have set $4\pi\alpha^{2}G=\beta$. Thus, in order to recover the Bekenstein-Hawking formula in the limit $\beta\rightarrow0$, we must set $b/\epsilon=2\pi$. We see from Eq.~(\ref{7}) that within the framework of the GUP the correction to the Bekenstein-Hawking formula for a spherically symmetric black hole is due to a multiplicative factor that depends on the area $A$ of the event horizon (see also \cite{Nasser, Ali1}.)
\section{Finding the $f(R)$-Modified Gravity}\label{sec:3}
Comparing formulas (\ref{3}) and (\ref{7}), we immediately see that both formulas imply, in accordance with the Bekenstein-Hawking formula, that the entropy is proportional to the area $A$ of the black hole's event horizon. The formulas differ, however, on the nature of the factor responsible for the departure from a pure area-law. This factor is, according to Wald's formula, entirely due, in an $f(R)$-gravity, to the constant curvature $R_{0}$ of the Universe. In the GUP approach, the departure from the pure area-law of the entropy is due, in an $f(R)$-gravity, to both $R_{0}$ \textit{and} the mass $M$ of the black hole. Indeed, it is known \cite{Cruz1, Cruz2} that for a black hole solution of mass $M$ in an $f(R)$-gravity, the metric in the coordinates $(t,r,\theta,\phi)$ has the form $\mathrm{d}s^{2}=-\lambda(r)\mathrm{d}t^{2}+\lambda^{-1}(r)\mathrm{d}r^{2}+r^{2}(\mathrm{d}\theta^{2}+\sin^{2}\theta\mathrm{d}\phi^{2})$, with $\lambda(r)=1-(2GM/r)-(R_{0}r^{2}/12)$. The black hole horizon $\mathcal{H}$ is then located at $r_{\mathcal{H}}$ for which $\lambda(r_{\mathcal{H}})=0$, and the area of the event horizon is given by $A=4\pi r^{2}_{\mathcal{H}}$. Therefore, when solving for $r_{\mathcal{H}}$, equation $\lambda(r_{\mathcal{H}})=0$ gives us actually $r_{\mathcal{H}}=r_{\mathcal{H}}(R_{0},M)$, and hence, $A=A(R_{0},M)$ instead of simply $A=A(R_{0})$, which could have been in perfect agreement with the entropy formula (\ref{3}) coming from Wald's approach.

To remove this conflict between the two approaches, we will look closer at each of the assumptions made in the previous section that led us to the result (\ref{7}). Looking again at the five steps we have used, we notice that except for the last two steps, all of the other steps were general and did not rely on any particular restrictive assumption and did not evoke the nature of the system involved. It is only at the fourth and fifth steps that we made specific choices concerning the thermodynamics of the system, and only there did we make contact with the quantum processes involved in generating the huge entropy of the black hole. In the fourth step we have assumed that the uncertainty $\Delta x$ is of the order of the characteristic scale of the system, which is the Schwarzschild diameter $2r_{S}$, whereas in the fifth step we have assumed that the elementary increase $\Delta S$ in entropy due to an elementary increase in the horizon area $\Delta A_{min}$ is constant. It is clear that only the fourth assumption is responsible for the form of entropy (\ref{7}) since the last assumption only fixes the multiplicative factor in the final result.

The choice of $\Delta x$, as we saw, was based on the fact that the black hole is in thermal equilibrium with the bath of particles whose wavelengths are of the order of the Hawking temperature. However, without being restrictive about the nature of the system, and regardless of the nature of the radiation coming out of the system, what was expected from the fourth step was simply to find the minimum increase $\Delta A_{min}$ of the horizon area after the absorption by the latter of a quantum particle. But we know, actually, that the black hole is absorbing other quanta besides Hawking quanta. In fact, the smallest increase of the black hole horizon might come from quanta of longer wavelengths thanks to the inequality $\Delta A_{min}\geq 8\pi l_{P}^{2}E\Delta x$. Such much smaller-energy quanta are naturally obtained from the geometry of the Universe that contains the black hole itself. Indeed, given that we took the curvature of the Universe to be the constant $R_{0}$, this gives us a radius of curvature of the order $\sim1/\sqrt{R_{0}}$. Therefore, we must take into account that the black hole is also absorbing quanta whose Compton wavelengths, or 'size', as seen by an asymptotic observer, are of the order $\Delta x\sim 1/\sqrt{R_{0}}$. Choosing this more 'generous' estimate for $\Delta x$ inside formula (\ref{5}) that gives the minimum bound for $\Delta A_{min}$, turns (\ref{6}) into the following differential equation:
\begin{equation}\label{8}
\frac{\mathrm{d}S}{\mathrm{d}A}=\frac{b}{\Delta A_{min}}=\frac{\alpha^{2}bR_{0}}{4\pi\epsilon\left(1-\sqrt{1-4\alpha^{2}GR_{0}}\right)},
\end{equation}
which gives, up to an integration constant, the following entropy:
\begin{equation}\label{9}
S=\frac{bA}{4\pi\epsilon}\left[\frac{\alpha^{2}R_{0}}{1-\sqrt{1-4\alpha^{2}GR_{0}}}\right].
\end{equation}
In order to recover the Bekenstein-Hawking formula in the limit $R_{0}\rightarrow0$, i.e. in the general relativity limit, we should set, as before, $b/\epsilon=2\pi$. We see that this last form looks similar to the one obtained from an $f(R)$-gravity via Wald's method. In fact, comparing the latter result with the formula (\ref{3}) obtained from Wald's approach, we can deduce at once the $f(R)$-gravity theory that could yield a black hole entropy which is in exact agreement with the one coming from the GUP. Equating the content of the square brackets in both formulas, with the value of $b/\epsilon$ in mind, and since the curvature constant $R_{0}$ we chose was arbitrary, we have the following differential equation for every $R$:
\begin{equation}\label{10}
\frac{\partial f(R)}{\partial R}=\frac{2\alpha^{2}GR}{1-\sqrt{1-4\alpha^{2}GR}},
\end{equation}
which can easily be integrated to give the following exact formula for $f(R)$:
\begin{equation}\label{11}
f(R)=\frac{R}{2}-\frac{(1-4\alpha^{2}GR)^{\frac{3}{2}}}{12\alpha^{2}G}+\mathrm{const.},
\end{equation}
where $\mathrm{const.}$ is the constant of integration. This last result does not look attractive for an $f(R)$-gravity theory yet, but a Taylor expansion in the Ricci scalar gives an expression that looks more familiar since it introduces various powers of the Ricci scalar:
\begin{equation}\label{12}
f(R)=R-\frac{\alpha^{2}GR^{2}}{2}-\Lambda+\mathcal{O}(R^{3}).
\end{equation}
Here we have merged the constant of integration found in (\ref{11}) with the constant $(-12\alpha^{2}G)^{-1}$, coming from the Taylor expansion, into the single constant $\Lambda$ which one might obviously want to identify with the cosmological constant. Power-law modified gravity models (as well as their recent variant \cite{Hammad2, Hammad3}) have been studied in the literature for their ability to unify both the early- and the late-time expansions of the Universe \cite{Nojiri1, Capozziello1, Felice, Capozziello2}.

Now, the $f(R)$-gravity we found in (\ref{11}) does not satisfy the constraint any cosmologically viable theory of modified gravity should satisfy, namely the inequalities $f_{R}>0$, $f_{RR}>0$ \cite{Felice}, and $L_{S}^{2}(f_{R}/f_{RR}-R)_{R=R_{0}}\gg1$ \cite{Olmo}, where $f_{R}$ and $f_{RR}$ represent, respectively, the first and second derivatives of $f(R)$ with respect to $R$ and $L_{S}$ is a length of the order of the solar system scale. Indeed, from (\ref{11}) we have $f_{RR}=-\alpha^{2}G/\sqrt{1-4\alpha^{2}GR}<0$.

Fortunately, though, there is yet another version of the GUP besides inequality (\ref{4}), already suggested in Ref.~\cite{Ali2}, which, in addition to the quadratic term in the uncertainty $\Delta p$ on the right-hand side, contains also a linear term in $\Delta p$ \cite{Ali1}:
\begin{equation}\label{13}
\Delta x\Delta p\geq\frac{1}{2}\left[1-\frac{4\sqrt{\mu}}{3}\alpha l_{P}\Delta p+2(1+\mu)\alpha^{2}l^{2}_{P}\Delta p^{2}\right],
\end{equation}
where $\alpha$, this time, is a dimensionless constant whose exact bound could only be fixed experimentally whereas $\mu$ is another dimensionless constant equal to $(2.82/\pi)^{2}$ \cite{Ali1}. Then, starting from this form of the GUP and repeating exactly the three steps that led us from (\ref{4}) to (\ref{5}) gives, at the leading order, the following minimum increase in area of the event horizon \cite{Ali1}:
\begin{equation}\label{14}
\Delta A_{min}\geq8\pi l^{2}_{P}\left(1-\frac{2\alpha l_{P}\sqrt{\mu}}{3\Delta x}\right).
\end{equation}
We see that if one uses again the estimate $\Delta x\sim2r_{S}=\sqrt{A/\pi}$, one finds from this result, after performing the fifth step that leads to the differential equation for entropy, that the latter would depart from a pure area-law by a multiplicative factor that depends on the horizon area $A$ \cite{Ali1}. Choosing $\Delta x\sim1/\sqrt{R_{0}}$, however, one finds the following differential equation instead:
\begin{equation}\label{15}
\frac{\mathrm{d}S}{\mathrm{d}A}=\frac{b}{\Delta A_{min}}=\frac{3b}{8\pi\epsilon G\left(3-2\alpha \sqrt{\mu GR_{0}}\right)},
\end{equation}
where we have introduced Newton's gravitational constant as well as the numerical factor $\epsilon$ that would fixe the minimum bound for $\Delta A_{min}$. We see that in order to recover the Bekenstein-Hawking result when $R_{0}\rightarrow0$, we should set again $b/\epsilon=2\pi$. Integrating Eq.~(\ref{15}) with this value for $b/\epsilon$ gives, up to an integration constant, the following entropy:
\begin{equation}\label{16}
S=\frac{A}{4G}\left[\frac{3}{3-2\alpha \sqrt{\mu GR_{0}}}\right].
\end{equation}
Comparing Eqs.~(\ref{3}) and (\ref{16}), we obtain the differential equation satisfied by $f(R)$ for every $R$, which gives after integration
\begin{equation}\label{17}
f(R)=-\frac{3}{\alpha}\sqrt{\frac{R}{\mu G}}-\frac{9}{2\mu\alpha^{2}G}\ln\left(3-2\alpha \sqrt{\mu GR}\right)+\mathrm{const}.
\end{equation}
This last result does not also look attractive for an $f(R)$-gravity theory yet, but a Taylor expansion in the Ricci scalar gives an expression that looks more familiar since it introduces various powers of the Ricci scalar:
\begin{equation}\label{18}
f(R)=R+\frac{4\alpha\sqrt{\mu G}}{9}R^{3/2}-\Lambda+\mathcal{O}(R^{2}),
\end{equation}
where we have again merged the constant of integration with the constant $-9\ln3/(2\mu\alpha^{2}G)$, coming from the Taylor expansion, into the single constant $\Lambda$ to evoke the cosmological constant. Now, unlike expression (\ref{11}), the modified gravity (\ref{17}) does not violate the above mentioned conditions that any modified gravity model should satisfy in order to be viable. Indeed, on one hand, we have from (\ref{17}) that $f_{R}>0$, $f_{RR}>0$. On the other hand, using the truncated expansion (\ref{18}), which is sufficient for our purposes here, we find $L_{S}^{2}(f_{R}/f_{RR}-R)_{R=R_{0}}\sim3L_{S}^{2}\sqrt{R_{0}}/(\alpha\sqrt{\mu G})\gg1$, where we have kept only the leading terms in the difference inside the parenthesis, and then used the fact that $R_{0}$ is of an order given by the observed Hubble parameter $\sim H_{0}^{2}$ and that $L_{S}$ is a length that ranges from meters to planetary scales \cite{Ali1}.

\section{Wald Entropy from GUP}\label{sec:4}
In the previous section we saw that two different versions of the GUP implied two different $f(R)$-modified gravity theories. Therefore, if one should rely on the GUP to find the unique modified gravity theory that would extend general relativity, one should, first of all, find, if any, the one and unique fundamental GUP.

The fact that the GUP, written in some versions, might be used to reproduce the same black hole entropy as the one predicted by Wald's entropy within a specific $f(R)$-gravity, and the fact that Wald's entropy works for any $f(R)$ theory, leads one to suspect that it could be possible to find a general version of the GUP which would yield for every modified gravity the same entropy as predicted by Wald's formula? For that to happen, however, one would have to find a formulation of the GUP such that the black hole entropy it reproduces for every $f(R)$-gravity theory would be nothing but the Wald formula (\ref{3}). In other words, we would like to investigate in this section the following question: Is it possible to find a GUP from which the fundamental Wald entropy formula would arise?

For that purpose, let us assume the following general form of the GUP
\begin{equation}\label{19}
\Delta x\geq h(\Delta p),
\end{equation}
where $h$ is a smooth and invertible function that could in principle be determined. The first step when using the GUP to extract the black hole entropy is, as explained in Sec.~\ref{sec:2}, to invert inequality (\ref{19}) and find the lower bound for the uncertainty on momentum $\Delta p$ in terms of the uncertainty on position $\Delta x$. Let us assume that the inverse function $h^{-1}$ is a decreasing function. Therefore, inverting inequality (\ref{19}) gives
\begin{equation}\label{20}
\Delta p\geq h^{-1}(\Delta x).
\end{equation}
The assumption of a decreasing function $h^{-1}$ implies physically that the system imposes a bigger uncertainty on momentum whenever we have a smaller uncertainty on position. This assumption, however, is not necessary. It simply allows one to extract (\ref{20}) from (\ref{19}). As we shall see below, one might instead have to start from (\ref{20}) without any assumption on $h^{-1}$ and deduce an equivalent of (\ref{19}). Before we proceed further, notice that for the special function $h^{-1}(y)=1/y$, inequality (\ref{20}) reduces to the usual Heisenberg uncertainty relation.

The next two steps consist in trading $\Delta p$ for $E$ and multiplying both sides of the resulting inequality by $8\pi l_{P}^{2}\Delta x$, to find $\Delta A_{min}\geq8\pi l_{P}^{2}\Delta xh^{-1}(\Delta x)$. The fourth step then relates the minimum increase in area to the constant Ricci scalar as, $\Delta A_{min}=8\pi\epsilon GR_{0}^{-\frac{1}{2}}h^{-1}(R_{0}^{-\frac{1}{2}})$. Finally, the last step gives rise to the following differential equation for entropy:
\begin{equation}\label{21}
\frac{\mathrm{d}S}{\mathrm{d}A}=\frac{b}{\Delta A_{min}}=\frac{1}{4G}\frac{R_{0}^{\frac{1}{2}}}{h^{-1}(R_{0}^{-\frac{1}{2}})},
\end{equation}
where we have relied on the Bekenstein-Hawking formula to set again $b/\epsilon=2\pi$, for the above differential equation must yield $S=A/4G$ when $h^{-1}(R^{-\frac{1}{2}})=R^{\frac{1}{2}}$ since for this special function, as we saw above, one recovers the usual Heisenberg uncertainty principle.

By integrating Eq.~(\ref{21}), and comparing the result with Wald's formula (\ref{3}), we deduce that we should have for every $R$
\begin{equation}\label{22}
h^{-1}(R^{-\frac{1}{2}})\frac{\partial f(R)}{\partial R}=R^{\frac{1}{2}}.
\end{equation}
Now this identity does not give us straightforwardly the original function $h$ appearing in inequality (\ref{19}) that gives the uncertainty on position in terms of the uncertainty on momentum, but the identity could be used in inequality (\ref{20}) to extract instead the uncertainty on momentum from that on position. Indeed, from (\ref{22}) we deduce that $h^{-1}(\Delta x)=1/[\Delta xf'(1/\Delta x^{2})]$, where the prime means a first derivative of the function. Finally, substituting this in inequality (\ref{20}), yields
\begin{equation}\label{23}
\Delta x\Delta p\geq\frac{1}{f'(\Delta x^{-2})}.
\end{equation}
This is a GUP from which Wald's general entropy formula for any diffeomorphism invariant $f(R)$-theory of gravity would result.

The physical meaning suggested by this inequality is that modified gravity, i.e. geometry, is behind the additional uncertainties beside the usual limitations found in quantum mechanics due there to the finite resolution of any measurement process. In fact, for the simple case of general relativity, whose Lagrangian is based on $f(R)=R$, (\ref{23}) reduces to the usual Heisenberg inequality. This does not however mean that general relativity does not imply any minimum length. The latter indeed appears at the level of horizons, since one already has $\Delta A_{min}\sim l_{P}^{2}$, regardless of the corrections brought to the Heisenberg relations.

Finally, we would also like to remark here that it is satisfactory that (\ref{23}) is physically consistent only for viable modified gravity models for which $f'>0$ \cite{Felice}.

\section{Conclusion and Discussion}
We have exposed in this paper two well-known derivations of the entropy of black holes. The first one is based on Wald's entropy formula and the second is based on the generalized uncertainty principle. Although the two methods are both fundamental, the result they give for the entropy of the same spherically symmetric black hole, taken from the vacuum solution of $f(R)$-gravity theories, is not the same. The difference lies in the corrections to the Bekenstein-Hawking formula the two approaches predict. Using Wald's formula, one finds that the deviation from a pure area-law is solely a function of the constant scalar curvature of the Universe containing the black hole. Using the GUP, on the other hand, one finds that those corrections should be a function of the area of the black hole's event horizon. Although the area of a black hole in an $f(R)$-gravity depends also on the scalar curvature of the Universe, the two results cannot be compatible since the corrections found within one approach would be different for different black holes whereas the corrections arising within the other approach would be the same for all black holes belonging to the same Universe.

This conflict between the two approaches becomes even more disturbing when we recall that the GUP came from fundamental approaches to quantum gravity that predict corrections to general relativity, whereas Wald's method, based on the fundamental concept of Noether charge, is valid for any diffeomorphism invariant theory of spacetime, i.e. it should be applied for whatever correction to general relativity one might consider.

Our proposal here to remove this conflict was to slightly modify the usual derivations of black holes's entropy by reconsidering the minimum area increase induced on an event horizon after the black hole absorbs a quantum of energy. By taking into account the absorption of all possible quanta within the Universe containing the black hole, and not just those whose wavelengths are of the order of the Hawking radiation, we found a new minimum of area increase of the horizon such that the entropy variation with area depends only on the scalar curvature of the Universe.

This reconciliation between the two approaches turned out to be an alternative way to search for $f(R)$-modified gravity theories. This is due to the fact that whereas the GUP formula gives the precise dependence on the curvature scalar of the multiplicative factor in the Bekenstein-Hawking formula, it does not tell us anything about the gravitational Lagrangian that caused this factor. In contrast, Wald's formula is capable of restituting back the analytic formula, i.e. the gravitational Lagrangian, behind any correction to the pure area-law. Therefore, the two approaches are complementary when it comes to searching for modified gravity theories.

Another very interesting feature of this procedure is that one version of the GUP formula, the one with more terms on the right-hand side of the inequality, gives a viable modified gravity model whereas the other version does not. So we expect that the more terms one includes inside the GUP formula, the closer one gets to a realistic $f(R)$-gravity, and more precise the model becomes since each additional term in the GUP comes with a factor of greater power of Newton's constant $G$.

In this paper, we restricted our study to the one-dimensional version of the GUP but one could also treat the full three-dimensional case \cite{Ali2}. The use of the latter would allow one to make contact with $f(R)$-gravity in an anisotropic spacetime \cite{Sharif,Saaidi}. Conversely, one could, as we saw in Sec.~\ref{sec:4} thanks to inequality (\ref{23}), also use one's favorite model of $f(R)$-modified gravity to constraint the GUP in the same spirit as done in Ref.~\cite{Gangopadhyaya}.

Finally, as we discussed in Sec.~\ref{sec:2}, we chose to examine $f(R)$-gravity for its formal simplicity as well as for the abundance of literature devoted to its study. Our approach could, however, readily be generalized to investigate more complicated gravitational Lagrangians since Wald's formula is valid for any diffeomorphism invariant theory.

\begin{acknowledgments}
I would like to thank Mir Faizal for having brought to my attention Ref.~\cite{Awad}. I would like also to thank the anonymous referee for his/her helpful comments.
\end{acknowledgments}

\end{document}